\pdfoutput=1
\documentclass[pra,twocolumn,superscriptaddress]{revtex4-2}
\usepackage{graphicx,empheq,amssymb,scalerel,tikz}
\usepackage[colorlinks=true, citecolor=blue, urlcolor=blue, linkcolor=blue ]{hyperref}
\usetikzlibrary{svg.path}
\definecolor{orcidlogocol}{HTML}{A6CE39}
\tikzset{
	orcidlogo/.pic={
		\fill[orcidlogocol] svg{M256,128c0,70.7-57.3,128-128,128C57.3,256,0,198.7,0,128C0,57.3,57.3,0,128,0C198.7,0,256,57.3,256,128z};
		\fill[white] svg{M86.3,186.2H70.9V79.1h15.4v48.4V186.2z}
		svg{M108.9,79.1h41.6c39.6,0,57,28.3,57,53.6c0,27.5-21.5,53.6-56.8,53.6h-41.8V79.1z M124.3,172.4h24.5c34.9,0,42.9-26.5,42.9-39.7c0-21.5-13.7-39.7-43.7-39.7h-23.7V172.4z}
		svg{M88.7,56.8c0,5.5-4.5,10.1-10.1,10.1c-5.6,0-10.1-4.6-10.1-10.1c0-5.6,4.5-10.1,10.1-10.1C84.2,46.7,88.7,51.3,88.7,56.8z};}}
\newcommand\orcid[1]{\href{https://orcid.org/#1}{\mbox{\scalerel*{\begin{tikzpicture}[yscale=-1,transform shape]\pic{orcidlogo};\end{tikzpicture}}{|}}}}

\begin{document}
\title{Hybrid-order topological odd-parity superconductors via Floquet engineering}
\author{Hong Wu\orcid{0000-0003-3276-7823}}
\affiliation{Key Laboratory of Quantum Theory and Applications of MoE, Lanzhou University, Lanzhou 730000, China}
\affiliation{Lanzhou Center for Theoretical Physics, Key Laboratory of Theoretical Physics of Gansu Province, Lanzhou University, Lanzhou 730000, China}
\author{Jun-Hong An\orcid{0000-0002-3475-0729}}\email{anjhong@lzu.edu.cn}
\affiliation{Key Laboratory of Quantum Theory and Applications of MoE, Lanzhou University, Lanzhou 730000, China}
\affiliation{Lanzhou Center for Theoretical Physics, Key Laboratory of Theoretical Physics of Gansu Province, Lanzhou University, Lanzhou 730000, China}
\begin{abstract}
Having the potential for performing quantum computation, topological superconductors have been generalized to the second-order case. The hybridization of different orders of topological superconductors is attractive because it facilitates the simultaneous utilization of their respective advantages. However, previous studies found that they cannot coexist in one system due to the constraint of symmetry. We propose a Floquet engineering scheme to generate two-dimensional (2D) hybrid-order topological superconductors in an odd-parity superconductor system. Exotic hybrid-order phases exhibiting the coexisting gapless chiral edge states and gapped Majorana corner states not only in two different quasienergy gaps but also in one single quasienergy gap are created by periodic driving. The generalization of this scheme to the 3D system allows us to discover a second-order Dirac superconductor featuring coexisting surface and hinge Majorana Fermi arcs. Our results break a fundamental limitation on topological phases and open a feasible avenue to realize topological superconductor phases without static analogs.
\end{abstract}
\maketitle
\section{Introduction}
As one of the central fields in modern physics, topological phases of matter not only enrich the paradigm of condensed matter physics and stimulate the discovery of many novel quantum materials, but also generate various fascinating applications in quantum technologies \cite{RevModPhys.83.1057,RevModPhys.88.021004,RevModPhys.88.035005,Sato_2017,RevModPhys.90.015001}. In the family of topological phases, the topological superconductor has attracted wide attention. It successfully simulates the mysterious Majorana fermion in a condensed matter system \cite{Wilczek2009}. It also may be a promising application in realizing quantum computation due to its non-Abelian statistics \cite{PhysRevLett.86.268,KITAEV20032,RevModPhys.80.1083}. Recently, there is an intense interest in extending the traditional topological phases to higher-order ones \cite{Benalcazar61,Schindlereaat0346,PhysRevLett.119.246402,PhysRevLett.118.076803}. Being parallel to the theoretical proposals \cite{PhysRevLett.124.166804,PhysRevLett.120.026801,PhysRevLett.126.206404,PhysRevLett.125.056402,PhysRevLett.124.036803,PhysRevLett.128.127601} and experimental observations \cite{PhysRevLett.122.233902,PhysRevLett.125.255502,PhysRevB.102.104109,PhysRevLett.125.213901,PhysRevLett.126.146802,PhysRevLett.122.204301,PhysRevLett.122.233903,PhysRevLett.126.156801,PhysRevX.11.011016,PhysRevB.104.224303} on second-order topological insulators, second-order topological superconductors (SOTSCs) have been proposed \cite{PhysRevLett.125.097001,PhysRevLett.121.096803,PhysRevB.97.205136,PhysRevLett.122.236401,PhysRevLett.119.246401,PhysRevLett.123.156801,PhysRevB.97.205135,PhysRevB.97.094508,PhysRevB.101.104502,PhysRevResearch.2.012060,PhysRevX.10.041014,PhysRevB.103.L041401,PhysRevB.106.125121}.
This opens another avenue toward topological quantum computation \cite{PhysRevB.97.205134,PhysRevB.100.045407,PhysRevResearch.2.032068,PhysRevResearch.2.043025}.

Explorations of physical systems supporting exotic topological features and of efficient ways to control these features are not only in the mainstream of condensed matter physics, but are also a demand of quantum technologies. Symmetry plays an important role in realizing SOTSCs \cite{PhysRevLett.125.097001,PhysRevB.97.205136,PhysRevLett.122.236401,PhysRevLett.119.246401,PhysRevLett.123.156801,PhysRevB.97.205135,PhysRevB.97.094508,PhysRevB.101.104502,PhysRevResearch.2.012060,PhysRevX.10.041014,PhysRevB.103.L041401,PhysRevB.106.125121}.
It is conventionally believed that SOTSCs are achieved by applying an even-parity term in an odd-parity superconductor to break the symmetry \cite{PhysRevB.99.020508,PhysRevB.97.205134,PhysRevB.98.165144,PhysRevLett.121.186801}. Recently, a general approach to realize SOTSCs in odd-parity superconductors was proposed \cite{PhysRevLett.123.177001}. An interesting question is whether the first-order and SOTSCs can coexist in one system. First, this kind of hybrid-order topological superconductor (HOTSC) facilitates the simultaneous utilization of the advantages of both corner states and gapless chiral edge states in designing bifunctional devices \cite{PhysRevLett.126.156801}. Second, it enriches the family of topological phases and leads to a different band theory. Nevertheless, according to the general view, first- and second-order topologies exist in different systems \cite{PhysRevLett.102.187001,PhysRevB.82.184516,PhysRevB.98.165144} due to their substantially different features in the energy spectrum at the boundaries \cite{PhysRevLett.102.187001,PhysRevB.82.184516,PhysRevLett.120.017001,PhysRevB.98.165144,PhysRevResearch.3.013239}. Therefore, it seems that a HOTSC is impossible to realize in a system under the framework of traditional topological energy-band theory.
On the other hand, coherent control via periodic driving, referred to as Floquet engineering, has become a versatile tool in artificially creating novel topological phases in systems of ultracold atoms \cite{RevModPhys.89.011004,PhysRevLett.116.205301}, photonics \cite{Rechtsman2013,PhysRevLett.122.173901}, superconductor qubits \cite{Roushan2017}, and graphene \cite{McIver2020}. Many exotic phases absent in static systems have been controllably generated by periodic driving \cite{PhysRevB.87.201109,PhysRevB.100.085138,PhysRevResearch.2.013124,PhysRevB.103.115308,PhysRevResearch.1.032013,PhysRevLett.124.216601,Jangjan2020,PhysRevResearch.1.032045,PhysRevB.103.045424,PhysRevB.103.085413,PhysRevB.105.L121113,PhysRevResearch.3.023039,PhysRevB.105.115418,PhysRevB.105.155406}.
One of the interesting findings is that periodic driving can cause different high-symmetry points to have different topological charges \cite{PhysRevB.93.184306}, which lays the foundation to realize different types of topological phases \cite{PhysRevB.104.205117}. Is it possible to realize HOTSCs by periodic driving?

Addressing these questions, we propose a general scheme to create a two-dimensional (2D) HOTSC in a two-band odd-parity superconductor system by Floquet engineering. A complete topological characterization to such a Floquet quasienergy band structure is established. We discover two kinds of HOTSCs, where in one of them the coexisting gapless chiral edge states and the gapped Majorana corner states reside in two different quasienergy gaps, and in the other they reside in one common gap. The generalization to the three-dimensional (3D) odd-parity superconductor system reveals another exotic phase, i.e., a second-order Dirac superconductor, which features coexisting surface and hinge Majorana Fermi arcs. Breaking the conventional constraint on realizing hybrid-order topology, our result reveals Floquet engineering as a useful way to explore different topological phases.

\section{Static system}
\begin{figure}[tbp]
\centering
\includegraphics[width=1\columnwidth]{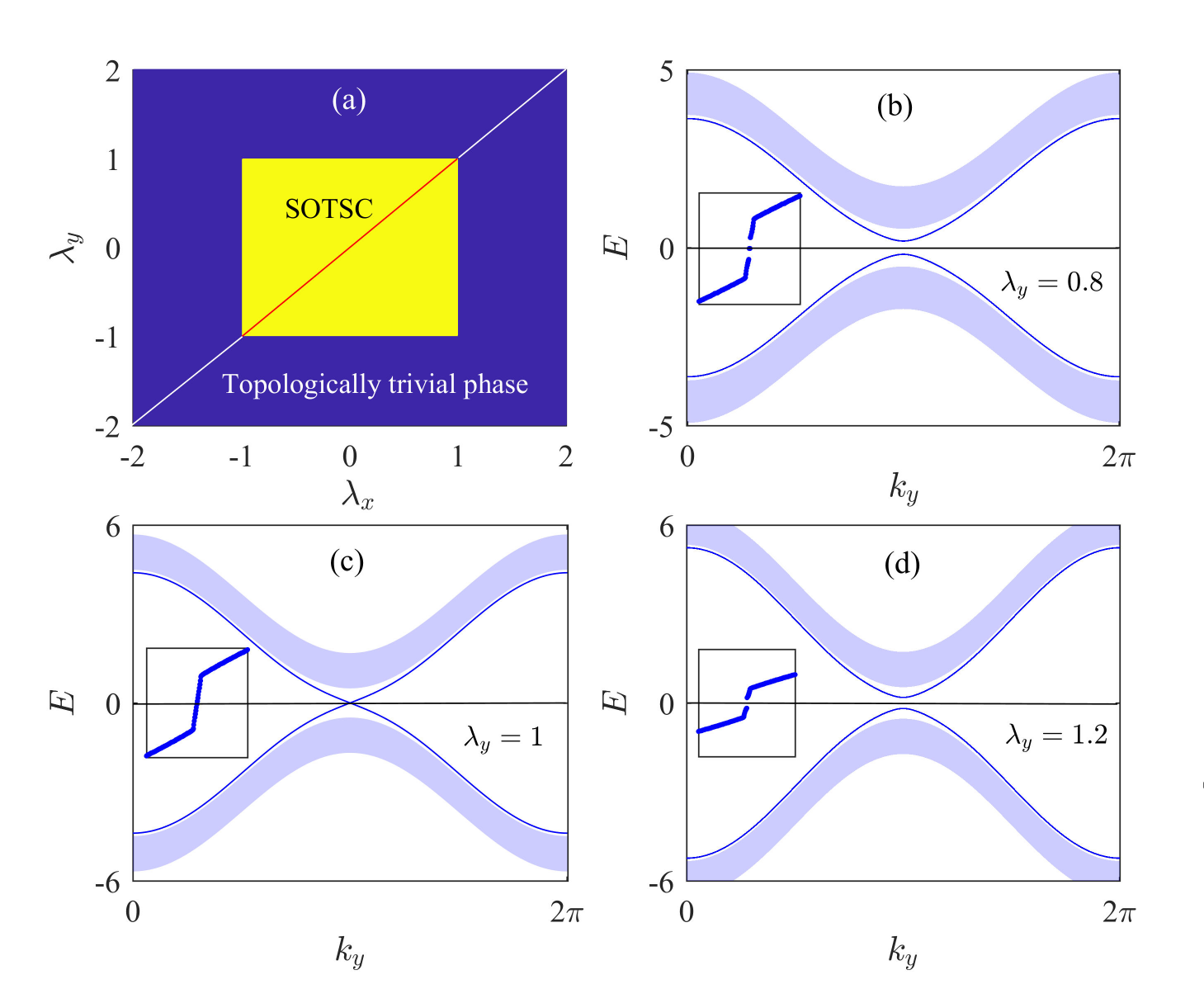}
\caption{(a) Phase diagram in the $\lambda_x$-$\lambda_y$ plane. The red and white lines mark the regimes with $\mathcal{W}=1$ and $0$, respectively. Energy spectra under the $x$-direction open boundary condition when (b) $\lambda_y=0.8$, (c) 1.0, and (d) 1.2. The inset in (a)-(c) shows the energy spectrum under the $x$- and $y$-direction open boundary condition. We use $\lambda_x=0.3$ and lattice sizes $L_x=L_y=50$.}  \label{static}
\end{figure}

Conventional studies suggested that superconductors with appropriate mixed-parity pairings are candidates for SOTSCs \cite{PhysRevB.99.020508,PhysRevB.98.165144}. Reference \cite{PhysRevLett.123.177001} presented a scheme to realize SOTSCs in purely odd-parity pairing superconductors. We use this static system as the starting point of our Floquet engineering. We consider a $p$-wave superconductor whose Hamiltonian reads $\hat{H}_1=\frac{1}{2}\sum_{\mathbf{k}}\hat{\Psi}^{\dag}_{\mathbf{k}} \mathcal{H}_1(\mathbf{k})\hat{\Psi}_{\mathbf{k}}$, with $\hat{\Psi}_{\mathbf{k}}^{\dag}=(\hat{c}_{\mathbf{k}}^{\dag},\hat{c}_{-\mathbf{k}})$ and
\begin{equation}
\mathcal{H}_1(\mathbf{k})=\left(
  \begin{array}{ccc}
    \epsilon(\mathbf{k}) & \Delta(\mathbf{k}) \\
    \Delta^*(\mathbf{k}) & -\epsilon(\mathbf{k}) \\
  \end{array}
\right),\label{sttc}
\end{equation}
where $\epsilon(\mathbf{k})$ is the kinetic energy of the normal state and $\Delta({\bf k})$ is the pairing potential of the superconductor. Equation \eqref{sttc} has spatial inversion symmetry $\mathcal{P}=\sigma_z$. We focus on the inversion symmetric normal state $\epsilon(\mathbf{k})$ and the odd-parity pairings satisfying $\Delta(\mathbf{k})=-\Delta(-\mathbf{k})$. Via a Hopf map \cite{PhysRevLett.118.147003}, a SOTSC model which is trivial in the first-order topology in this odd-parity superconductor is constructed as $\Delta({\bf k})=2(\cos k_x+\lambda_x+i\cos k_y+i\lambda_y)(\sin k_x-i\sin k_y)$ and $\epsilon({\bf k})=\sum_{j=x,y}[(\cos k_j+\lambda_j)^2-\sin^2k_j]$. The SOTSC is formed when $|\lambda_{x,y}|<1$, where four gapped zero-mode Majorana corner states are formed [see Figs. \ref{static}(a) and \ref{static}(b)]. Different from conventional topological phases, this SOTSC is not caused by the closing and reopening of the bulk-band gap. The energy spectra under the $x$-direction open boundary condition in Figs. \ref{static}(b)-\ref{static}(d) indicate that, when the parameter runs from the SOTSC to the topologically trivial regimes, the bulk-band gap is persistently open, however, the boundary-state gap occurs a closing and reopening. Therefore, we cannot find a topological invariant of the bulk bands to characterize the SOTSC. This type of SOTSC is referred to as the boundary obstructed topological phase \cite{PhysRevResearch.3.013239}. A counterexample arises when $\lambda_x=\lambda_y$, where the system has a mirror-rotation symmetry $\sigma_y\mathcal{H}_1(k_x,k_y)\sigma_y=-\mathcal{H}_1(k_y,k_x)$. Its topology can be described by the mirror-graded winding number defined in the bulk bands as $\mathcal{W}=\frac{i}{2\pi}\int_{0}^{2\pi}{\langle u(k)\lvert\partial_k\lvert u(k)\rangle}dk$, where $\lvert u(k)\rangle$ is the eigenstate of $\mathcal{H}_1({k})$, along the high-symmetry line $k_x=k_y\equiv k$ [see Fig. \ref{static}(a)].

\section{Floquet engineering}
To generate HOTSC, we apply a periodic driving
\begin{equation}
\mathcal{H}({\bf k},t)=\mathcal{H}_1({\bf k})+\mathcal{H}_2({\bf k})\delta(t/T-n),\label{odr}
\end{equation}
where $\mathcal{H}_2=m\sigma_z$, $T$ is the driving period, and $n$ is an integer. The periodic system $\mathcal{H}({\bf k},t) = \mathcal{H}({\bf k},t + T)$ does not have an energy spectrum because its energy is not conserved. According to the Floquet theorem, the one-period evolution operator $U(T)=\mathbb{T}\exp[-i\int_{0}^{T}\mathcal{H}({\bf k},\tau)d\tau]$ defines $\mathcal{H}_\text{eff}({\bf k})\equiv iT^{-1}\ln U(T)$ whose eigenvalues are called quasienergies \cite{PhysRevB.93.184306}. The topological phases of our system are defined in such a quasienergy spectrum. Applying the Floquet theorem to Eq. \eqref{odr}, we have $\mathcal{H}_\text{eff}({\bf k})=iT^{-1}\ln [e^{-i\mathcal{H}_2({\bf k})T}e^{-i\mathcal{H}_1({\bf k})T}]$ \cite{PhysRevLett.113.236803}. Different from the static case, our topological phase transition occurs not only at zero quasienergy but also at $\pi/T$ due to the periodicity of the quasienergies, which may cause the Floquet anomaly in pure-order topological phases \cite{PhysRevX.3.031005}.

We first focus on Floquet engineering when $\lambda_x=\lambda_y\equiv\lambda$, where the phases are characterized by the bulk bands. First, we can derive from $\mathcal{H}_\text{eff}({\bf k})$ that the phase transition occurs when the parameters satisfy either
\begin{eqnarray}
&2(1-\lambda^2)-m=n\pi/T& \label{nlp1}\\
&\text{or}~~2[1+\lambda^2+\lambda(e^{i\alpha_x}+e^{i\alpha_y})]+m=n_{\alpha_x,\alpha_y}\pi/T& \label{npcd2}
\end{eqnarray}
at the quasienergy zero ($\pi/T$) for even (odd) $n$ and $n_{\alpha_x,\alpha_y}$, with $\alpha_{x,y}=0$ or $\pi$, (see Appendix \ref{dadadd}). Giving the phase boundaries of our system, Eqs. \eqref{nlp1} and \eqref{npcd2} offer us sufficient space to artificially synthesize exotic topological phases absent in the static system \eqref{sttc} by periodic driving. Second, we can establish a complete topological characterization on $\mathcal{H}_\text{eff}({\bf k})$ to the rich emergent topological phases in our periodic system. Besides the intrinsic SOTSC in the static case, our periodic driving also induces a first-order Chern superconductor phase. We have to find appropriate topological invariants to characterize the first- and second-order superconductor phases occurring at both zero and $\pi/T$ quasienergies. The overall topology of the first- and second-order phases at the $\alpha/T$ quasienergy, with $\alpha=0$ or $\pi$, is described by $\mathcal{W}_{\alpha/T}=(\mathcal{W}_1+e^{i\alpha}\mathcal{W}_2)/2$, where $\mathcal{W}_j$ are the mirror-graded winding numbers defined in $\tilde{\mathcal{H}}_j({\bf k})=iT^{-1}\ln [G_j U(T)G_j^\dag]$, with $G_j=e^{i(-1)^j\mathcal{H}_j({\bf k})T/2}$, along the high-symmetry line $k_x=k_y=k$ \cite{PhysRevB.90.125143}. The first-order topology at the $\alpha/T$ quasienergy is described the dynamical winding number \cite{PhysRevB.104.L180303,PhysRevB.96.155118}
\begin{eqnarray}
\mathcal{V}_{\alpha/ T}=\int_{0}^{T}dt\frac{d^2{\bf k}}{24\pi^2}\epsilon_{ijk} {\rm Tr}(\mathcal{U}^{\dag}_{\alpha}\partial_{i} \mathcal{U}_{\alpha} \mathcal{U}^{\dag}_{\alpha}\partial_{j} \mathcal{U}_{\alpha} \mathcal{U}^{\dag}_{\alpha}\partial_{k} \mathcal{U}_{\alpha}).~~
\end{eqnarray}
Here $\epsilon_{ijk}$, with $\{i,j,k\}\in \{t, k_x,k_y\}$, is the completely antisymmetric tensor and a sum over repeated indices has been used, $\mathcal{U}_{\alpha}=U(t)\sum_{l=1}^2e^{i\mathcal{E}^{(\alpha)}_{l,{\bf k}}t}|u_{l}({\bf k},T)\rangle\langle u_{l}({\bf k},T)|$, where $\mathcal{E}^{(\alpha)}_{l,\mathbf{k}}$ is the quasienergy in the $l$th band chosen in the regime $[\alpha/T,(\alpha +2\pi)/T)$ and $|u_{l}({\bf k},T)\rangle$ is the corresponding Floquet eigenstate. The number of the first-order gapless edge states and the second-order gapped Majorana corner states at the $\alpha/T$ quasienergy is equal to $|\mathcal{V}_{\alpha/T}|$ pairs and $2(|\mathcal{W}_{\alpha/T}|-|\mathcal{V}_{\alpha/T}|)$, respectively.

\begin{figure}[tbp]
\centering
\includegraphics[width=1\columnwidth]{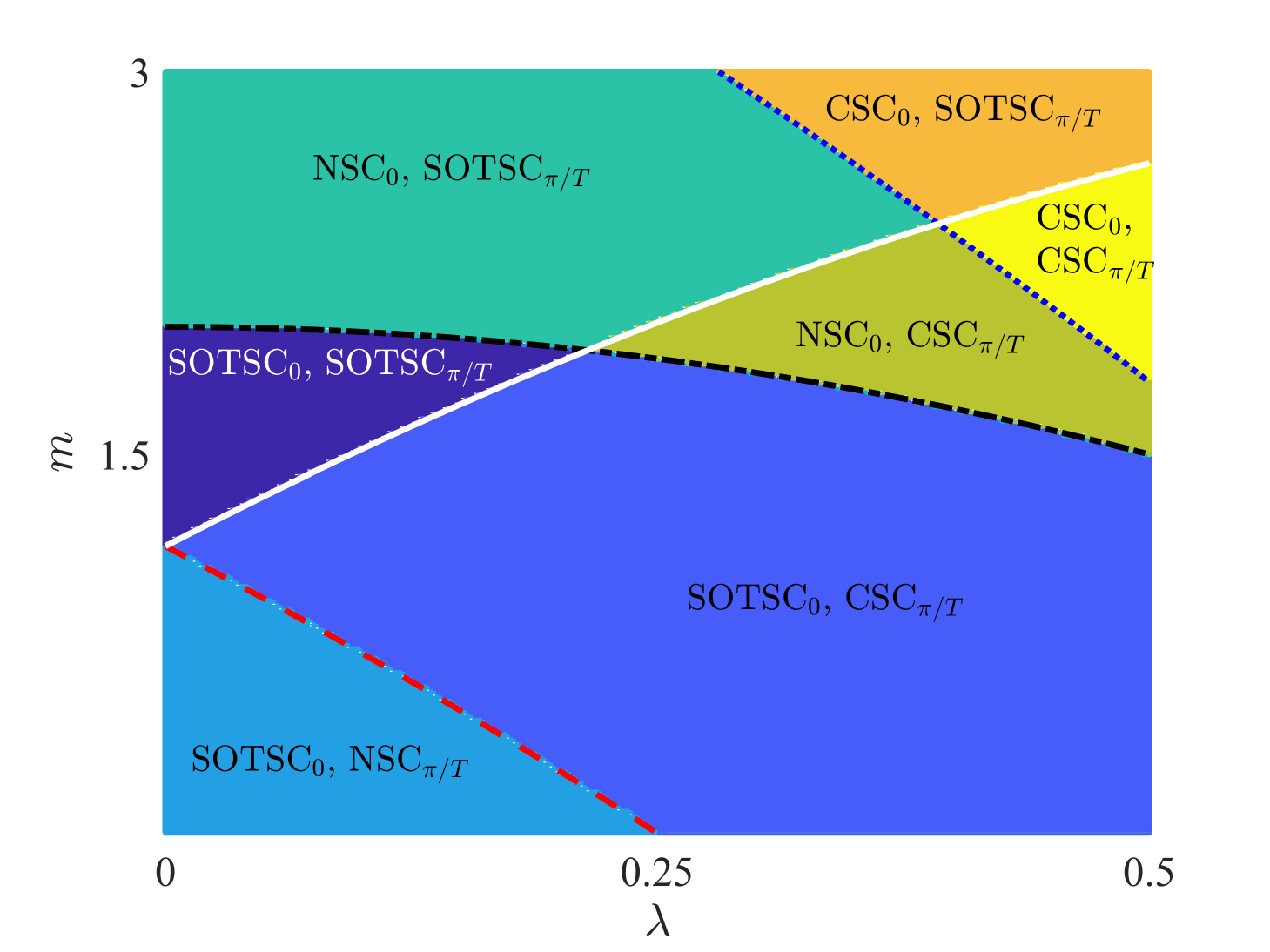}
\caption{Phase diagram in the $\lambda$-$m$ plane characterized by $\mathcal{W}_{\alpha/T}$ and $\mathcal{V}_{\alpha/T}$ ($\alpha=0,\pi$). The black dotted-dashed line is from Eq. \eqref{nlp1} with $n=0$. The  red dashed, blue dotted, and white solid lines are from Eq. \eqref{npcd2} with $n_{0,0}=1$, $2$, and $n_{\pi,\pi}=1$, respectively. NSC$_{\alpha/T}$ denotes the topologically trivial normal superconductor with $\mathcal{W}_{\alpha/T}=\mathcal{V}_{\alpha/T}=0$, CSC$_{\alpha/T}$ denotes the first-order Chern superconductors with $|\mathcal{W}_{\alpha/T}|=1$ and $|\mathcal{V}_{\alpha/T}|=1$, and SOTSC$_{\alpha/T}$ is the SOTSC with $|\mathcal{W}_{\alpha/T}|=2$ and $|\mathcal{V}_{\alpha/T}|=0$. We use $T=1$.} \label{traj}
\end{figure}

To reveal the topological characters of our periodic system, we plot in Fig. \ref{traj} the phase diagram characterized by $\mathcal{W}_{\alpha/T}$ and $\mathcal{V}_{\alpha/T}$ ($\alpha=0,\pi$). We find that the system exhibits more colorful topological phases than the static one in Fig. \ref{static}(a). Its topological features are not only carried by the quasienergy gaps at both zero and $\pi/T$, but also become diverse for a given gap. We discover that rich phases, which may be any possible pair combination of the topologically trivial normal superconductor, first-order Chern topological superconductor, and SOTSC coexisting in the two quasienergy gaps, are created by periodic driving. Each phase boundary perfectly obeys either Eq. \eqref{nlp1} or \eqref{npcd2}. The result indicates that, permitting us to realize diverse exotic HOTSCs absent in its static counterpart, periodic driving supplies us with another dimension to change the topology orders and realize different topological phases at will.

\begin{figure}[tbp]
\centering
\includegraphics[width=1\columnwidth]{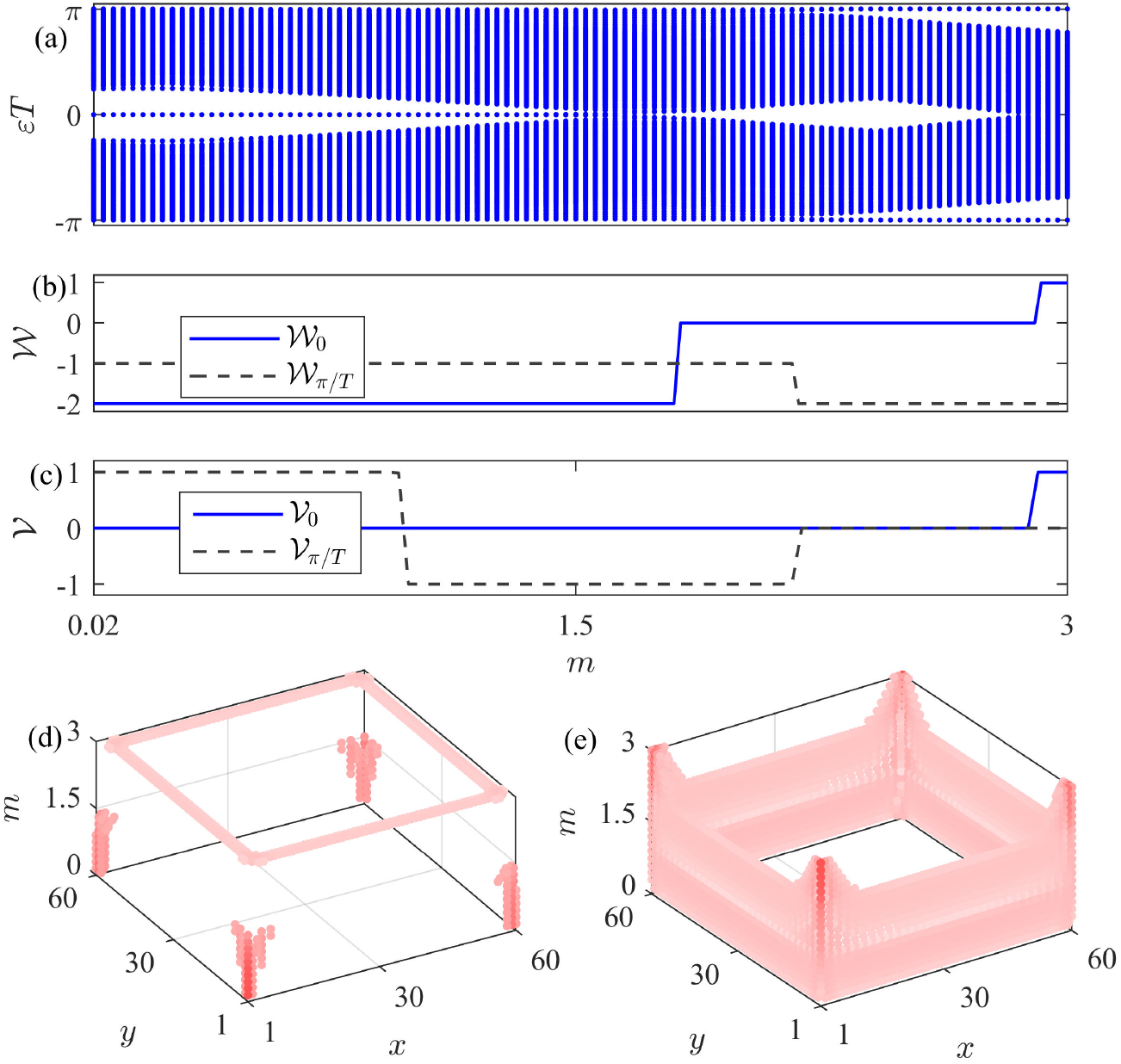}
\caption{(a) Quasienergy spectrum, (b) $\mathcal{W}_{\alpha/T}$, (c) $\mathcal{V}_{\alpha/T}$, and probability distributions of the (d) zero- and $\pi/T$-mode states in different $m$. We use $T=1$ and $\lambda=0.3$.} \label{5}
\end{figure}

Figure \ref{5}(a) shows the quasienergy spectrum in different $m$ for a given $\lambda$. We observe three typical regimes from the zero quasienergy, i.e., the regimes with gapped zero-mode states when $m<1.82$, with an open zero-mode gap when $1.82<m<2.9$, and with a closed gap when $m\geq2.9$. The topological invariants $\mathcal{W}_0$ and $\mathcal{V}_0$ in Figs. \ref{5}(b) and \ref{5}(c) reveal that these regimes correspond exactly to the SOTSC with four Majorana corner states, the normal superconductor, and the first-order Chern superconductor with one pair of gapless edge states, respectively. At the $\pi/T$ quasienergy, the gap is closed when $m\leq2.16$ and the gapped $\pi/T$-mode states are formed when $m>2.16$. The two regimes, according to $\mathcal{W}_{\pi/T}$ and $\mathcal{V}_{\pi/T}$ in Figs. \ref{5}(b) and \ref{5}(c), correspond to the first-order Chern superconductor and SOTSC, respectively. The features in this quasienergy spectrum match well with the phase diagram in Fig. \ref{traj}. We show the probability distributions of the zero- and $\pi/T$-mode states in Figs. \ref{5}(d) and \ref{5}(e). A clear coexistence of the first-order chiral edge and the second-order Majorana corner states at the zero and $\pi/T$ quasienergies consistent with Fig. \ref{5}(a) is found. It confirms the hybrid-order nature of the topological odd-parity superconductor phases generated by our periodic driving protocol.

\begin{figure}[tbp]
\centering
\includegraphics[width=1\columnwidth]{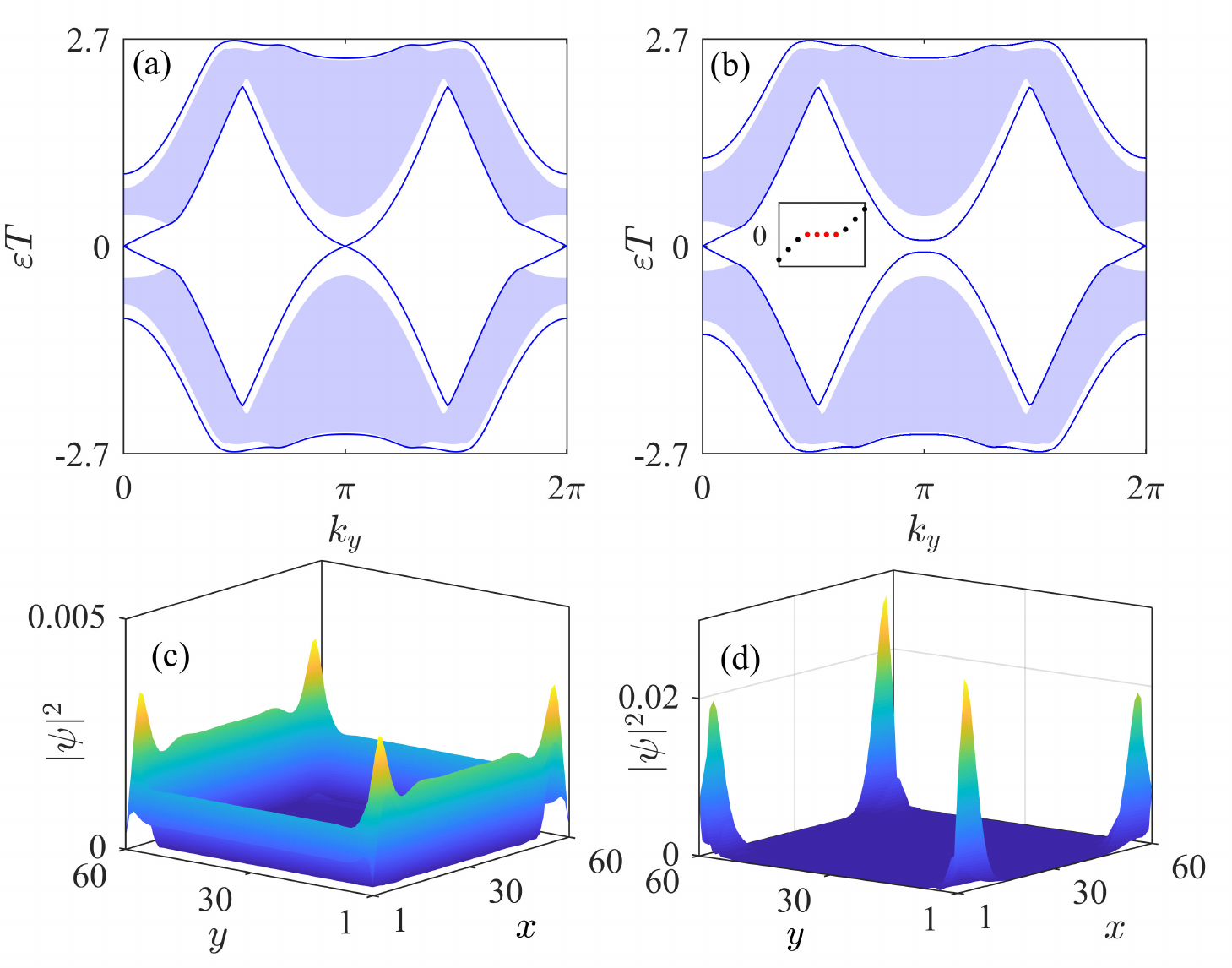}
\caption{Quasienergy spectra under the $x$-direction open boundary condition when (a) $\lambda_y=1$ and (b) $0.95$ (b). The inset of (b) shows the quasienergies under the $x$- and $y$-direction open boundary condition. (c) Probability distributions of the gapless chiral zero-mode edge state and (d) the gapped Majorana corner state near the zero quasienergy when $\lambda_y=0.95$. We use $\lambda_x=0.3$, $m=0.6$, and $T=1.1$.} \label{4}
\end{figure}

A generalization of our result to the case of $\lambda_x\neq \lambda_y$ can be made. An exotic HOTSC in one single quasienergy gap is realizable. Applying periodic driving on the critical phase in Fig. \ref{static}(c), we find in Fig. \ref{4}(a) that two extra gapless chiral edge states are present near $k_y=0$ in the quasienergy gap of zero. By decreasing the system parameter $\lambda_y$, the boundary-state gap at $k_y=\pi$ is opened and four gapped boundary states are formed [see Fig. \ref{4}(b)]. The probability distributions of the gapless edge state at $k_y=0$ in Fig. \ref{4}(c) and the gapped boundary state at $k_y=\pi$ in Fig. \ref{4}(d) confirm that they are first-order edge and second-order Majorana corner states, respectively. The evidence proves that it is a HOTSC with coexisting SOTSC and Chern superconductors in the same quasienergy gap. Although it is not protected by bulk topology, it also exhibits a certain robustness to the static and temporal disorders due to the inversion symmetry (see Appendix \ref{dssordreapp}).

\section{3D second-order Dirac superconductor}
\begin{figure}[tbp]
\centering
\includegraphics[width=1\columnwidth]{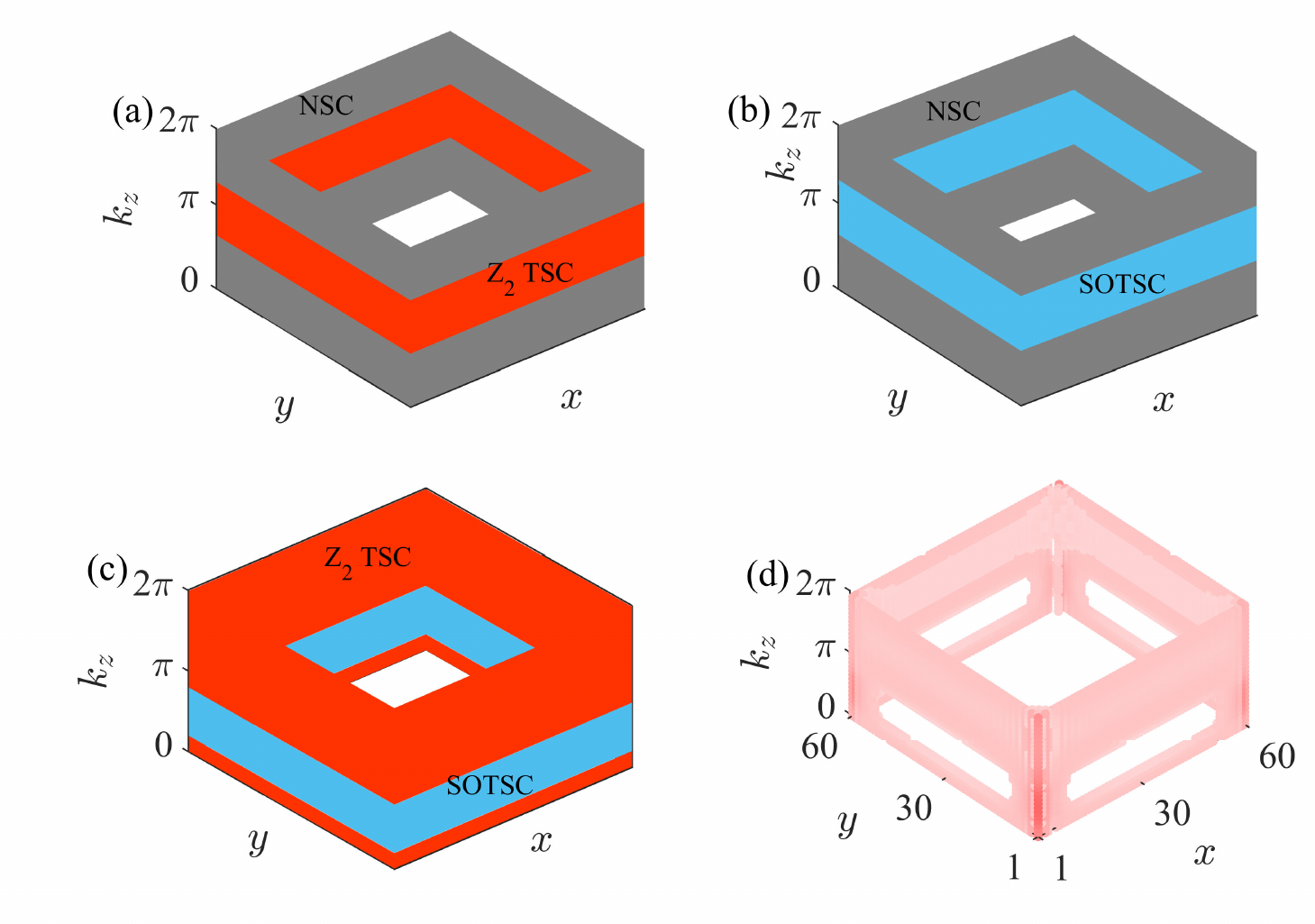}
\caption{ Schematic illustration of conventional (a) first- and (b) second-order and (c) our second-order Dirac superconductors. The phase transitions between the 2D sliced $Z_2$ topological superconductor ($Z_2$ TSC) and the SOTSC occur at certain $k_z$. (d) Numerical result of the coexisting surface and hinge Majorana Fermi arcs contributed by the $\pi/T$-mode first-order gapless chiral edge  and second-order gapped Majorana corner states. We use $\lambda_x=\lambda_y=0.3$, $m=1.87$, $t_z=0.45$, and $T=1$.} \label{other}
\end{figure}

We generalize our result to the 3D odd-parity superconductor system with a spin degree of freedom. We consider the same driving protocol as Eq. \eqref{odr} but with $\mathcal{H}_1({\bf k})=d_xs_z\sigma_x+d_y\sigma_y+d_z\sigma_z$ and $\mathcal{H}_2({\bf k})=(t_z\sin k_z+m)\sigma_z$, where $s_{i}$ and $\sigma_i$ are the Pauli matrices, $d_x+id_y=\Delta({\bf k})$, and $d_z=\epsilon({\bf k})$. It is found that $\mathcal{H}_1({\bf k})$ and $\mathcal{H}_2({\bf k})$ are $\mathcal{PT}$ invariant, with $\mathcal{T}=is_2\sigma_0\mathcal{K}$ and $\mathcal{P}=s_0\sigma_z$ being the time-reversal and spatial-inversion operations, which are inherited by $\mathcal{H}_\text{eff}({\bf k})$. The 3D phase is sliced into a family of 2D $k_z$-parametrized phases. If they are a SOTSC or a normal superconductor for all $k_z$, then the 3D phase is SOTSC or normal superconductor. If a 2D phase transition occurs at certain $k_z$, then the 3D phase is a Dirac superconductor \cite{PhysRevB.100.020509,PhysRevB.102.094503}. The Dirac superconductor is conventionally classified into first and second order [see Figs. \ref{other}(a) and \ref{other}(b)]. Here, we can realize a distinct second-order Dirac superconductor.  Being independent of $k_z$, $\mathcal{H}_1({\bf k})$ possesses the same energy spectrum as the 2D case in Eq. \eqref{sttc}. Thus, the static phase of $\mathcal{H}_1({\bf k})$ is a 3D SOTSC or normal superconductor depending on the parameters in Fig. \ref{static}(a). Using the parameter of $\mathcal{H}_1({\bf k})$ in the SOTSC regime and switching on the periodic driving, it is remarkable to find that a 3D second-order Dirac superconductor is realized [see Fig. \ref{other}(c)]. Such a unique topological phase is featured with coexisting surface and hinge Majorana Fermi arcs [see Fig. \ref{other}(d)]. Although a similar coexistence has been found in Weyl semimetals \cite{PhysRevLett.125.146401,PhysRevLett.125.266804}, it has yet to be found in Dirac semimetals and superconductors. It reveals that Floquet engineering offers us a useful tool to manipulate the types of Majorana Fermi arcs in $\mathcal{PT}$-symmetric systems.

\section{Discussion and conclusion}
It is noted that the delta-function driving protocol is considered just for the convenience of analytical calculation. Our scheme is generalizable to other driving forms. Our HOTSC phases are generalizable to other static systems constructed by the Hopf map \cite{PhysRevLett.123.177001}, which reveals the universality of our result. Boundary obstructed topological high-$T_c$ superconductivity is predicted in Ca$_{1-x}$La$_{x}$FeAs$_{2}$ \cite{PhysRevX.10.041014}. Floquet engineering has been used to design novel topological phases in several platforms \cite{RevModPhys.89.011004,PhysRevLett.116.205301,Rechtsman2013,PhysRevLett.122.173901,Roushan2017,McIver2020}. The higher-order topological phases have been simulated in Josephson junction platforms \cite{PhysRevB.98.104515,PhysRevLett.120.067001} and circuit QED systems \cite{NIU20211168}, where in the latter our static model has been realized. Based on these developments, we believe that our proposal can be realized in state-of-the-art quantum simulation platforms.

In summary, we have investigated the periodic-driving-induced topological phase transition in a two-band odd-parity superconductor system. A general Floquet engineering scheme to create exotic HOTSCs exhibiting coexisting first-order chiral edge and second-order Majorana corner states not only in two different quasienergy gaps but also in one single quasienergy gap has been established. The generalization of this scheme to a 3D system in the presence of $\mathcal{PT}$ symmetries permits us to realize a second-order Dirac superconductor. This phase features coexisting surface and hinge Majorana Fermi arcs. Our result indicates that periodic driving offers us a feasible way to explore exotic topological phases by adding time periodicity as an extra control dimension in the system. This significantly expands the scope of topological materials and enriches their controllability.

\section*{Acknowledgment}
The work is supported by the National Natural Science Foundation (Grants No. 12275109, No. 12247101, and No. 11834005) and the Supercomputing Center of
Lanzhou University.

\appendix

\section{Boundaries of topological phase transition}\label{dadadd}
Applying the Floquet theorem to our periodically driven system $\mathcal{H}({\bf k},t)=\mathcal{H}_1({\bf k})+\mathcal{H}_2({\bf k})\delta(t/T-N)$, we obtain
\begin{equation}
\mathcal{H}_\text{eff}({\bf k})=iT^{-1}\ln [e^{-i\mathcal{H}_2({\bf k})T}e^{-i\mathcal{H}_1({\bf k})T}].
\end{equation}
Using $\mathcal{H}_j({\bf k})=\mathbf{h}_j\cdot{\pmb{\sigma}}$ and the Euler's formula of the Pauli matrices
$e^{-i\mathbf{h}_j\cdot\pmb{\sigma}T}=\cos(h_jT)-i\sin(h_jT)\underline{\mathbf{h}}_j\cdot\pmb{\sigma}$,
with $h_j=|\mathbf{h}_j|$ and $\underline{\mathbf{h}}_j=\mathbf{h}_j/h_j$, we have
\begin{eqnarray}
{U}_T&=&e^{-i\mathbf{h}_2\cdot\pmb{\sigma} T}e^{-i\mathbf{h}_1\cdot\pmb{\sigma} T}=\epsilon I_{2\times2}+i\mathbf{r}\cdot\pmb{\sigma}\nonumber\equiv e^{-i\mathcal{H}_\text{eff}T},\label{utt}\\
\epsilon&=&\cos(h_1T)\cos(h_2T)-\underline{\mathbf{h}}_1\cdot\underline{\mathbf{h}}_2\sin(h_1T)\sin(h_2T),\label{epsl}~~~~\\
 \mathbf{r}&=& \underline{\mathbf{h}}_1\times\underline{\mathbf{h}}_2\sin(h_1 T)\sin(h_2 T)-\underline{\mathbf{h}}_2\cos(h_1 T)\nonumber\\ &&\times\sin(h_2 T)-\underline{\mathbf{h}}_1\sin(h_1T)\cos(h_2T).\label{varepsilonds}
\end{eqnarray}
Again using the Euler's formula, we can infer $\mathcal{H}_\text{eff}$ from Eq. \eqref{utt} as
\begin{eqnarray}
\mathcal{H}_\text{eff}({\bf k})&=&-\arccos(\epsilon)\mathbf{r}\cdot\pmb{\sigma}/[T\sin(\arccos\epsilon)]\nonumber\\
&=&-\arccos(\epsilon)\mathbf{r}\cdot\pmb{\sigma}/[T\sqrt{1-\epsilon^2}]\nonumber\\
&=&-\arccos(\epsilon)\underline{\mathbf{r}}\cdot\pmb{\sigma}/T,
\end{eqnarray}
where $\epsilon^2+|\mathbf{r}|^2=1$ has been used. Thus we obtain $\mathbf{h}_\text{eff}=-{\arccos \epsilon\over T} \underline{\mathbf{r}}$. The quasienergy bands touch at $0$ and $\pi/T$, which occurs when $\epsilon=+1$ and $-1$, respectively. Therefore, the phase transition is associated with the closing of the quasienergy bands, which occurs for ${ k}$ and driving parameters satisfying any one of the following conditions.
\begin{enumerate}
  \item $\sin(h_1 T)\sin(h_2 T)=0$: In this case, $\epsilon=\cos(h_1 T)\cos(h_2 T)$. Then the bands of $\mathcal{H}_\text{eff}({\bf k})$ close when
  \begin{equation}
h_j T=n_j\pi,~n_j\in \mathbb{Z}. \label{gen}
  \end{equation}
  \item $\underline{\mathbf{h}}_1\cdot\underline{\mathbf{h}}_2=\pm1$: In this case, $\epsilon=\cos(h_1T\pm h_2T)$. Then the bands of $\mathcal{H}^\alpha_\text{eff}({\bf k})$ close when
  \begin{equation}
h_1 T\pm h_2 T=n\pi,~n\in\mathbb{Z},\label{hh1}
  \end{equation}
 at the zero quasienergy (or $\pi/T$) if $n$ is even (or odd).
\end{enumerate}
As a sufficient condition for the topological phase transition, Eqs. \eqref{gen} and \eqref{hh1} supply a guideline to manipulate the driving parameters for Floquet engineering to various topological phases at will.

\begin{figure}[tbp]
\centering
\includegraphics[width=1\columnwidth]{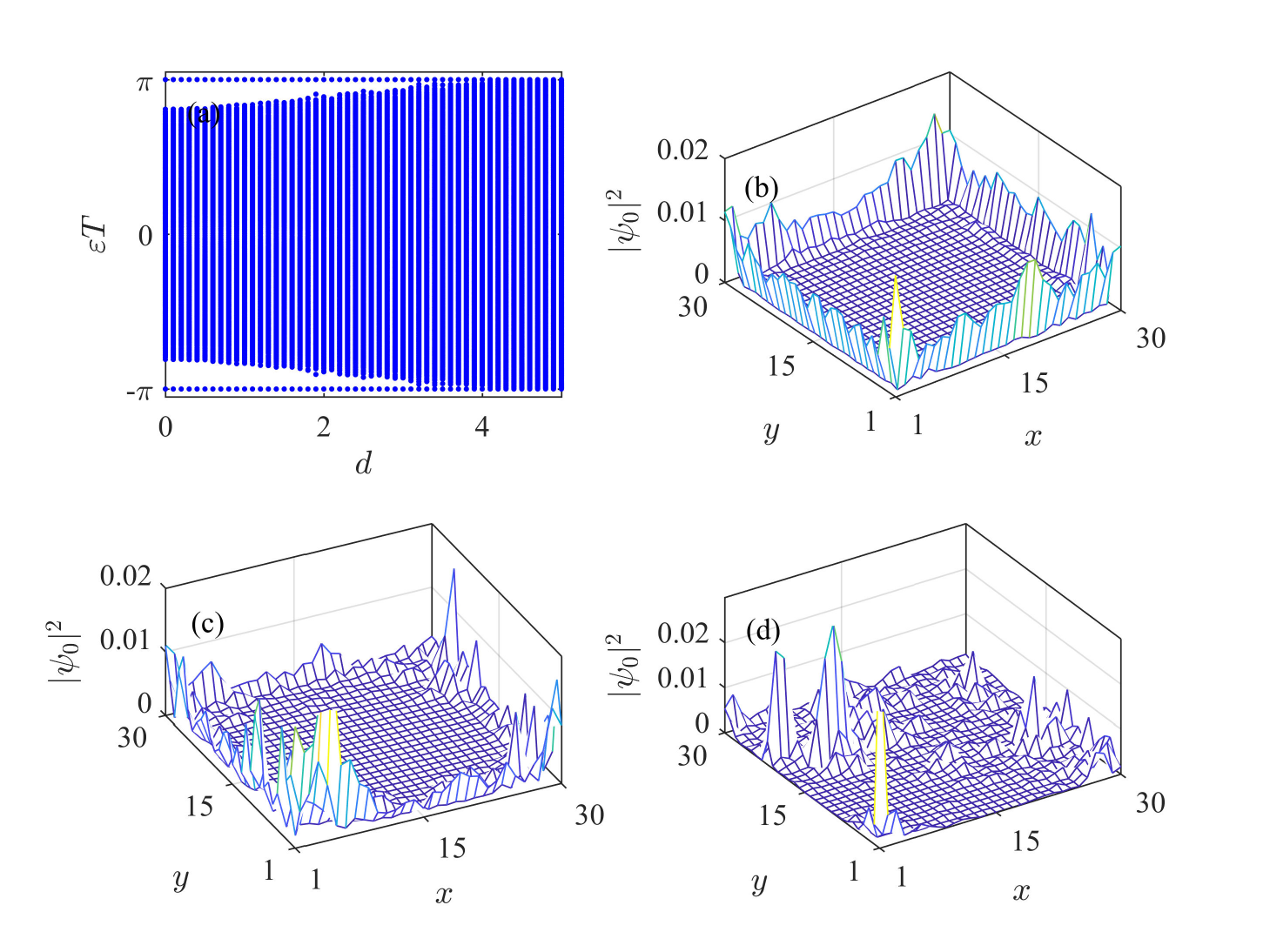}
\caption{(a) The quasienergy spectrum as a function of the disorder strength $d$ when a static disorder is considered. Probability distributions of the gapless zero-mode state when (b) $d=1$, (c) $2$, and (d) $3$. We use $\lambda_x=0.3$, $\lambda_y=0.4$, $m=0.38$, and $T=1$.} \label{traj1}
\end{figure}
Remembering $\mathcal{H}_2({\bf k})=m\sigma_z$ and rewriting $\mathcal{H}_1({\bf k})={\bf h}_1({\bf k})\cdot\pmb{\sigma}$ with ${\bf h}_1({\bf k})\equiv \text{Tr}[\mathcal{H}_1({\bf k})\pmb{\sigma}]$, we have
\begin{eqnarray}
h_{1}^x({\bf k})&=&2\sum_{j=x,y}(\cos k_j+\lambda)\sin k_j,\\
h_{1}^y({\bf k})&=&2[(\cos k_x+\lambda)\sin k_y-(\cos k_y+\lambda)\sin k_x],~~~\\
h_{1}^z({\bf k})&=&\sum_{j=x,y}[(\cos k_j+\lambda)^2-\sin^2k_j],\\
h_{2}^x({\bf k})&=&h_{2}^y({\bf k})=0,~h_{2}^z({\bf k})=m.
\end{eqnarray}
It can be checked that the band touching points given by Eq. \eqref{gen} generally do not cause a topological phase transition in the phase diagram \cite{PhysRevB.93.184306}. Then, according to the band touching condition in Eq. \eqref{hh1}, we obtain the boundaries of topological phase transition as follows.
\\ \textbf{Case I:} $\underline{\mathbf{h}}_1\cdot\underline{\mathbf{h}}_2=-1$ needs $\cos k_x=\cos k_y=-\lambda$. Then Eq. \eqref{hh1} requires
\begin{equation}
2(1-\lambda^2)T-mT=n\pi. \label{smnlp}
\end{equation}
\textbf{Case II:} $\underline{\mathbf{h}}_1\cdot\underline{\mathbf{h}}_2=1$ needs $k_{x,y}\equiv\alpha_{x,y}=0$ or $\pi$. Then Eq. \eqref{hh1} requires
\begin{equation}
2[1+\lambda^2+\lambda(e^{i\alpha_x}+e^{i\alpha_y})]T+mT=n_{\alpha_x,\alpha_y}\pi. \label{smnpcd}
\end{equation}

\begin{figure}[tbp]
\centering
\includegraphics[width=1\columnwidth]{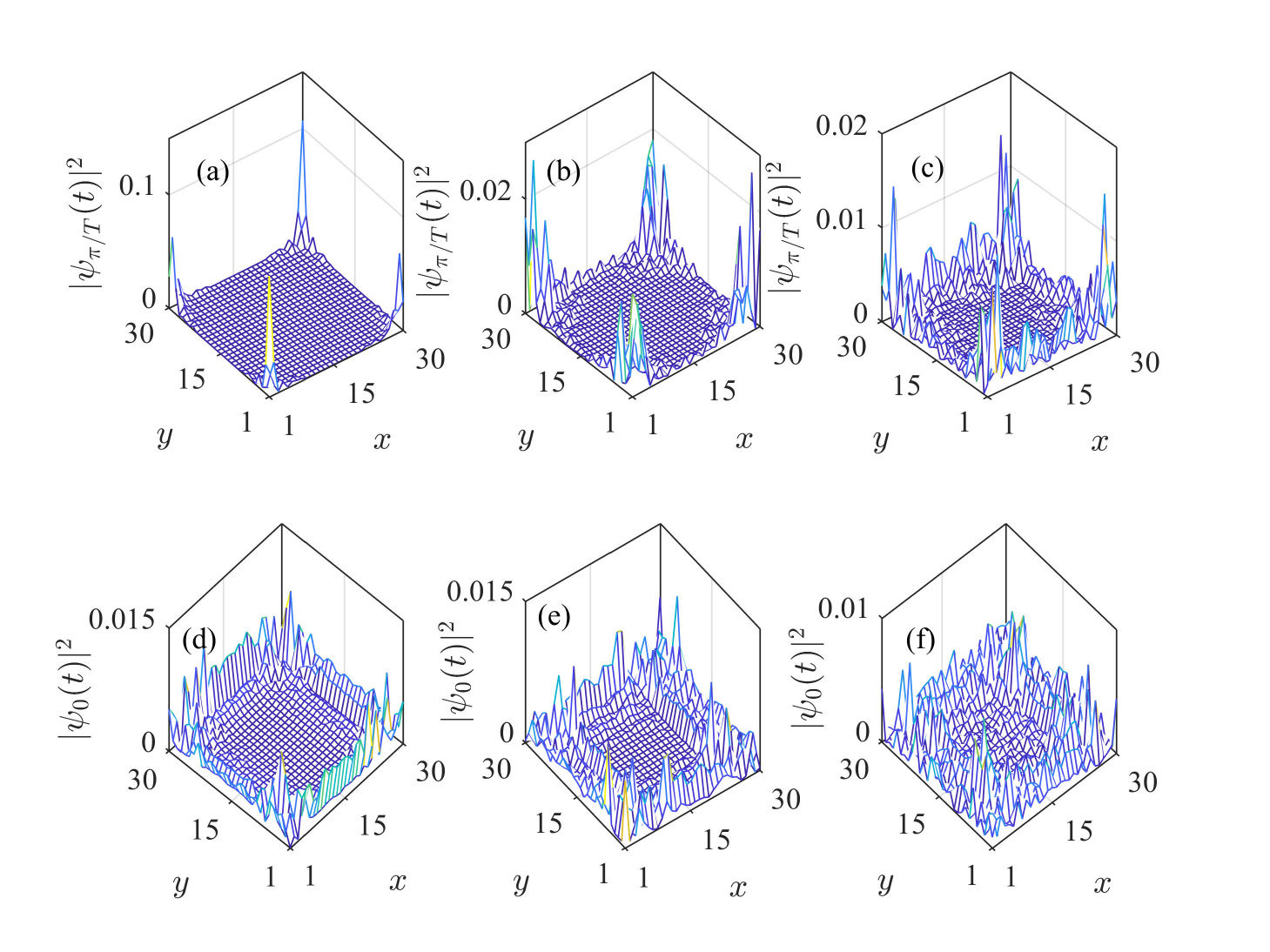}
\caption{When a temporal disorder is considered, the evolution of the probability distributions of the $\pi/T$-mode corner state at (a) $t=10T$, (b) $15T$, (c) $20T$, and the gapless zero-mode edge state (d) $t=10T$, (e) $20T$, (f) $30T$. We use $\lambda_x=0.3$, $\lambda_y=0.4$, $m=0.38$, and $T=1$.} \label{td}
\end{figure}
\section{Robustness to disorder}\label{dssordreapp}
We now show the robustness of our hybrid-order topological superconductor to the disorders.
According to Ref. \cite{PhysRevB.96.195303}, the existence of class A Floquet topological phases with gapless edge states does not require any symmetry. The higher-order topological odd-parity superconductors in our Floquet systems are protected by inversion symmetry. So our hybrid-order topological odd-parity superconductors are robust against disorders that preserve inversion symmetry.

After adding a disorder $d\xi$, where $\xi$ is a random number in the regime $[-0.5,0.5]$ and $d$ is the disorder strength, on the on-site potential $m$, we plot in Fig. \ref{traj1}(a) the quasienergy spectrum as a function of the disorder strength $d$. In the clean case when $d=0$, the system is a hybrid-order topological superconductor phase with a first-order gapless edge state in the zero mode and the second-order corner state in the $\pi/T$ mode. After turning on the disorder, we see that the $\pi/T$-mode corner state survives even when $d$ is as large as $4$. The probability distribution of the zero-mode edge state in Figs. \ref{traj1}(b)-\ref{traj1}(d) reveals that the zero-mode gapless edge state is present even when $d$ is as large as 2. With further increasing $d$, the two gaps close and the hybrid-order topological phase vanishes.

To reveal the robustness of our hybrid-order phase to the temporal disorder, we add a disorder $\mathcal{T}\xi$, where $\xi$ again is a random number in the regime $[-0.5,0.5]$ and $\mathcal{T}$ is the disorder strength, on the driving period $T$. It is naturally expected that the zero and the $\pi/T$-mode states would no longer be in the stationary state as long as $\mathcal{T}\xi$ is added on the driving period $T$. Therefore, they are not dynamically stable. Figures \ref{td}(a)-\ref{td}(c) show the evolution of the $\pi/T$-mode corner state when the disorder strength is $\mathcal{T}=0.5$. It is seen that the corner feature is stably presented even when the evolution time is as large as $20T$. Similar behavior is observed in the gapless zero-mode edge state [see Figs. \ref{td}(d)-\ref{td}(f)].

Therefore, our hybrid-order topological phases exhibit robustness to both of the static and temporal disorders.

\bibliography{references}
\end{document}